\documentstyle[12pt]{article}

\newcommand{\be}{\begin{equation}}
\newcommand{\ee}{\end{equation}}
\newcommand{\ba}{\begin{eqnarray}}
\newcommand{\ea}{\end{eqnarray}}
\newcommand{\ol}{\overline}
\newcommand{\ra}{\rightarrow}
\newcommand{\lb}[1]{\label{#1}}
\newcommand{\bb}[1]{\bibitem{#1}}

\begin{document}

\title{Flat wormholes from straight cosmic strings}
\author{G\'erard Cl\'ement
\thanks{E-mail: GECL@CCR.JUSSIEU.FR}
\\
\small Laboratoire de Gravitation et Cosmologie Relativistes\\ 
\small Universit\'e Pierre et Marie Curie, CNRS/URA769 \\
\small Tour 22-12, Bo\^{\i}te 142,
 4 place Jussieu, 75252 Paris cedex 05, France}
\date{July 4, 1996}
\maketitle
\abstract{
Special multi-cosmic string metrics are analytically extended to describe
configurations of Wheeler-Misner wormholes and ordinary cosmic strings. I
investigate in detail the case of flat, asymptotically Minkowskian,
Wheeler-Misner wormhole spacetimes generated by two cosmic strings, each
with tension $-1/4G$.
}

\section{Introduction}

The wormholes I shall discuss are traversable, Lorentzian wormholes
\cite{MT}. These are best defined by a simple example. Remove from
Euclidean space $R^3$ a volume $\Omega$. Take a second, identical copy of
$R^3 - \Omega$, and either identify these two excised spaces along the
boundaries $\partial\Omega$, or connect them by a cylindrical tube
$\Gamma\times\partial\Omega$. The resulting spatial geometry is a model of
an Einstein-Rosen \cite{ER} traversable wormhole, with two asymptotically
flat regions (actually, these are everywhere flat except on
$\partial\Omega$). One may similarly construct a model of a Wheeler-Misner
\cite{WM} wormhole, with only one asymptotically flat region, by removing
from $R^3$ two non-overlapping ``identical'' volumes $\Omega$ and
$\Omega'$, and connecting the boundaries $\partial\Omega$ and
$\partial\Omega'$ by a cylindrical tube. It has been speculated that
travelling through such hypothetical wormholes could shorten intergalactic
journeys.

Visser \cite{Visser} discussed the special case where $\Omega$ is a
polyhedron, and showed that in this case the curvature of the boundaries 
$\partial\Omega$ is
concentrated on the edges. The resulting Visser wormholes are thus generated
by arrays of cosmic strings (the polyhedra).
In this talk
I shall consider the case of parallel cosmic strings, following an
analytical approach complementary to the geometrical approach just
outlined. I shall first show how special multi-cosmic string metrics may
be analytically extended to Einstein-Rosen or Wheeler-Misner
multi-wormhole, multi-cosmic string metrics. Then I shall focus on the
special cases of Wheeler-Misner wormholes generated by two cosmic strings,
or by a single cosmic string. 

\setcounter{equation}{0}
\section{Wormholes from cosmic strings}

The well-known multi-cosmic string metric is
\be \lb{cosm}
ds^2 = dt^2 - d\sigma^2 - dz^2,
\ee
where the 2-metric
\be 
d\sigma^2 = \prod_i |\zeta - a_i|^{-8Gm_i} \,d\zeta \,d\zeta^\ast
\ee
($\zeta \equiv x + iy$) is everywhere flat:
\be \lb{flat}
d\sigma^2 = dw\,dw^\ast = du^2 + dv^2\,,
\ee
except for conical singularities (branch points of $\zeta(w)$) with
deficit angles $8\pi Gm_i$, located at the points $\zeta = a_i$.

Consider the special case of the bicone with $m_1 = m_2 = 1/8G$,
\be \lb{bicone}
d\sigma^2 = \frac{d\zeta \,d\zeta^\ast}{|\zeta^2 - b^2|}\,.
\ee
This may be analytically extended to a geodesically
complete surface: a cylinder. To show this \cite{worm}, pinch the cylinder
along a parallel. We thus obtain two identical bicones with deficit angles
$\pi$ at the two vertices, joined along the pinch. 
These two bicones are diffeomorphic to the two sheets of the Riemann
surface of the metric (\ref{bicone}) cut along the segment connecting the
two branch points. The diffeomorphism is implemented by the transformation
\be \lb{cosh}
\zeta = b \cosh w\,,
\ee
leading back to the cylinder metric (\ref{flat}) ($v = {\rm Im}\,w$ is 
an angular variable from (\ref{cosh})).

The cylinder with its two circles at infinity is the basic building block
for Einstein-Rosen wormholes. The metric for a flat space-time with  
$n$ wormholes and $2p$ ordinary cosmic strings is (\ref{cosm}) with
\be \lb{multiworm}
d\sigma^2 = \frac{\prod_{i=1}^p |\zeta - c_i|^{-8Gm_i}}{|\zeta_n^2 - b^
{2n}|} \,d\zeta \,d\zeta^\ast,
\qquad \zeta_n = \prod_{j=1}^n (\zeta - a_j)\,,
\ee
analytically extended to the Riemann surface made of two sheets joined
along the $n$-component cut. The deficit angle at infinity corresponds to
a total mass per unit length
\be \lb{M}
M = \frac{n}{4G} + \sum_{i=1}^p m_i\,.
\ee

In the case $n = 2$, $p = 0$, Eq. (\ref{M}) gives $M = 1/2G$, corresponding to
compact spatial two-sections. These being also regular and orientable can
only be tori $S^1 \times S^1$.
To recover the symmetrical Riemann surface, 
pinch the torus along two opposite circles; this yields two tetracones
with deficit angles $\pi$ at each vertex, joined along the two pinches,
which correspond to the two cuts of the Riemann surface. The
transformation from the tetracone metric to the torus metric is 
(in the case $a_1 + a_2 = 0$)
\be
\zeta = c\;{\rm sn}(cw, k)
\ee
($c$ and $k$ constant), where sn is a bi-periodical Jacobi function.

However the metric (\ref{multiworm}) with $n = 2$, $p = 0$ admits a more 
economical analytical
extension to a topologically non trivial Riemann surface with only one
sheet. The torus may be pinched only once into a single tetracone joined 
to itself by
an identification of the two edges. This identification corresponds to an
identification of the two cuts, leading to the identification $\zeta \ra
-\zeta$ of the two
sheets of the Riemann surface for the complex variable $\zeta(w)$. 
Such a one-sheeted extension is possible whenever the distribution of both
the $n$ cuts and the $p$ conical singularities of the metric (\ref{multiworm})
is invariant under the isometry $\zeta \rightarrow - \zeta$, so that the
two sheets of the symmetrical extension may be identified together. In the
case $n = 2$ the resulting surface ---a topological torus with a single 
point at
infinity and $p$ conical singularities--- is a flat Wheeler-Misner wormhole.
This may be asymptotically Minkowskian if $M = 0$, {\em i.e.}
$\sum_{i=1}^pm_i = - 1/2G$.

\setcounter{equation}{0}
\section{Two-string and one-string wormholes}   
The asymptotically Minkowskian Wheeler-Misner wormhole generated by two 
cosmic strings ($n = p = 2$) with negative mass/tension $m_1 = m_2 =
-1/4G$ is the one-sheeted extension of the metric 
\be \lb{2cs}
d\sigma^2 = \frac{|\zeta^2 - c^2|^2}{|(\zeta^2 - a^2)^2 -
b^4|}\,d\zeta\,d\zeta^{\ast}. 
\ee 
Depending on the values of the parameters $a$, $b$, $c$, we obtain two
basic geometries, DD or Q:

1) {\em DD wormhole} (``dipole-dipole''). The geodesic pattern leads to the
following geometrical construction of the $t=$ const., $z=$ const.
sections $\Sigma$. Remove from the Euclidean $(u, v)$ plane a rectangular 
strip, and glue together two opposite edges of length $2d$ of the
resulting boundary. Then glue the remaining two edges of length $2l$ to 
the two ends of a truncated cylinder of circumference $2l$ and length $2L$.

We define a ``path through the wormhole'' as a non--contractible closed path
in $\ol{\Sigma}$. There are here two such kinds of paths  corresponding to
the two circles of $S^1 \times S^1$. Paths crossing once the two
identified segments of length $2d$ are ``shorter'' (than they would be in
Euclidean space) by $2l$. Paths crossing once the two circular junctions
of length $2l$ may be ``shorter'' if $L < d$, but are always ``longer'' if
$L > d$. 

2) {\em Q wormhole} (``quadrupole''). This two-dimensional geometry may be
obtained by incising the Euclidean plane along a segment of length $2d$,
also incising a torus of circumferences $2l$ and $2L$ along a 
matching segment of a small circle ($2l > 2d$), and gluing together the
torus and the plane along the two edges of the cuts. In this case, paths
``through the wormhole'' (paths crossing once the two junctions) are always 
``longer'' by at least $2L$, and may be arbitrarily long, due to the
possibility of multiple windings around the large circle.

One-cosmic string wormholes may be obtained from the preceding geometries 
by taking limits such that the two cosmic strings (conical singularities)
coincide. There are again two basic geometries. The {\em DD$_0$ wormhole} 
is obtained by gluing together, first two opposite edges, then the other
two edges, of a rectangular hole in the Euclidean plane. To construct 
the {\em 8 wormhole}, incise the plane along
a segment, bring together the two vertices so that the
two edges make a figure 8, and glue these two edges to the two ends of
a truncated cylinder. These two geometries have a common limit, which
corresponds simply to a plane with two points identified.

In the geometrical optics approximation, light is not scattered by these
flat, asymptotically Minkowskian wormholes. However, light rays following
inequivalent paths may be shifted relative to each other. This effect,
which is similar to that  of a parallel plate, multiple reflections being 
replaced by multiple turns around the cylinder (or the torus), gives rise to a
one--dimensional (for the {\em DD} or {\em 8} wormholes), or two--dimensional 
(for the Q wormhole) array of images of a point source.
A fuller wave--optics treatment is under progress; one expects a
non--trivial effect arising both from diffraction by the topological
defects (cosmic strings) and resonance due to periodicity conditions in the
cylinder (or torus).

\end{document}